\documentclass[aps,reprint,showpacs,twocolumn,floatfix]{revtex4-1}

\usepackage{epsfig}
\usepackage{amsmath}
\usepackage{amssymb}
\usepackage{amsfonts}
\usepackage{mathptmx}
\usepackage{eucal}
\usepackage{bm}
\usepackage{color}
\usepackage[colorlinks,linkcolor=blue,citecolor=blue]{hyperref}
\usepackage{epstopdf}
\usepackage{graphicx}
\usepackage{natbib}
\begin{document}



\title{Rectification of electronic heat current by a hybrid thermal diode}

\author{M. J. Mart\'inez-P\'erez}
\affiliation{NEST, Istituto Nanoscienze-CNR and Scuola Normale Superiore, I-56127 Pisa, Italy}

\author{A. Fornieri}
\affiliation{NEST, Istituto Nanoscienze-CNR and Scuola Normale Superiore, I-56127 Pisa, Italy}

\author{F. Giazotto}
\email{f.giazotto@sns.it}
\affiliation{NEST, Istituto Nanoscienze-CNR and Scuola Normale Superiore, I-56127 Pisa, Italy}








\maketitle


\textbf{
Thermal diodes \cite{Starr,CasatiPRL}, i.e., devices allowing heat to flow preferentially in one direction, constitute one of the key tools for the implementation of solid-state thermal circuits. These would find application in many fields of nanoscience, e.g., cooling, energy harvesting, thermal isolation, radiation detection \cite{GiazottoRev}, quantum information \cite{NielsenChuang}, or emerging fields such as phononics \cite{LiRev,Dubi,WangLi} and coherent caloritronics \cite{GiazottoNature,MartinezNature,MartinezRev}. Yet, both in terms of \textit{phononic} \cite{Segal,LiAPL,LiPRL} or \textit{electronic} heat conduction \cite{Kuo}, which is the scope of this work, their experimental realization remains still very challenging \cite{RobertsRev}.
A highly-efficient thermal diode should provide differences of at least one order of magnitude between the heat current transmitted in the forward temperature ($T$)-bias configuration, $J_{fw}$, and that generated upon $T$-bias reversal, $J_{rev}$, leading to $\mathcal{R}=J_{fw}/J_{rev}\gg1$ or $\ll1$.
So far, $\mathcal{R}\sim1.07-1.4$ has been reported in \textit{phononic} devices \cite{SciRep,Chang,Kobayashi} whereas $\mathcal{R} \sim1.1$ was obtained with a quantum-dot \textit{electronic} thermal rectifier at cryogenic temperatures \cite{Scheibner}. Here we show that unprecedented ratios reaching $\mathcal{R}\sim140$ can be attained in a hybrid device combining normal metals tunnel-coupled to superconductors \cite{MartinezAPLrect,GiazottoBergeret,FornieriAPL}. Our approach provides with a high-performance realization of a thermal diode for the electronic heat current that could be successfully implemented in true low-temperature solid-state thermal circuits.}

As recently proposed, substantial rectification of the electronic heat current can be achieved in metallic microcircuits based on tunnel junctions at low temperatures \cite{MartinezAPLrect,GiazottoBergeret,FornieriAPL}. These simple elements, based on widespread fabrication technology and well-known physics, should indeed allow the realization of efficient electronic thermal diodes and have still to be realized experimentally. 
Two kinds of devices have been analyzed theoretically so far. One consisted of a NIS junction - where N stands for a normal metal, I for a thin insulating layer and S denotes a superconducting electrode - in which thermal symmetry is broken by the $T$-dependence of the energy gap ($\Delta$) in the superconducting density of states (DOS). $\mathcal{R}$ up to $\sim 0.8 $ was predicted to occur at temperatures close to the critical temperature ($T_{\rm c}$) of S \cite{MartinezAPLrect,GiazottoBergeret}. 
Strongly improved results were foreseen for a N$_{\rm L}$ININ$_{\rm R}$ chain - where subscripts L and R refer to the left and right leads, respectively - subjected to the following two conditions: first, the thermal coupling between the normal metal electrodes in the left junction ($\propto 1/R_{\rm L}$, $R_{\rm L}$ being the normal-state resistance of the N$_{\rm L}$IN contact) must differ largely from that in the right ($\propto 1/R_{\rm R}$); 
second, electrons in the central lead have to be coupled to the phonon bath.
Experimentally, this latter condition can be realized through a thermalizing cold finger tunnel-coupled to the N electrode. Under these circumstances, $\mathcal{R}$ up to $\sim 2000 $ can be theoretically achieved \cite{FornieriAPL}.

\begin{figure*}[t]
\centering
\includegraphics[width=0.65\textwidth]{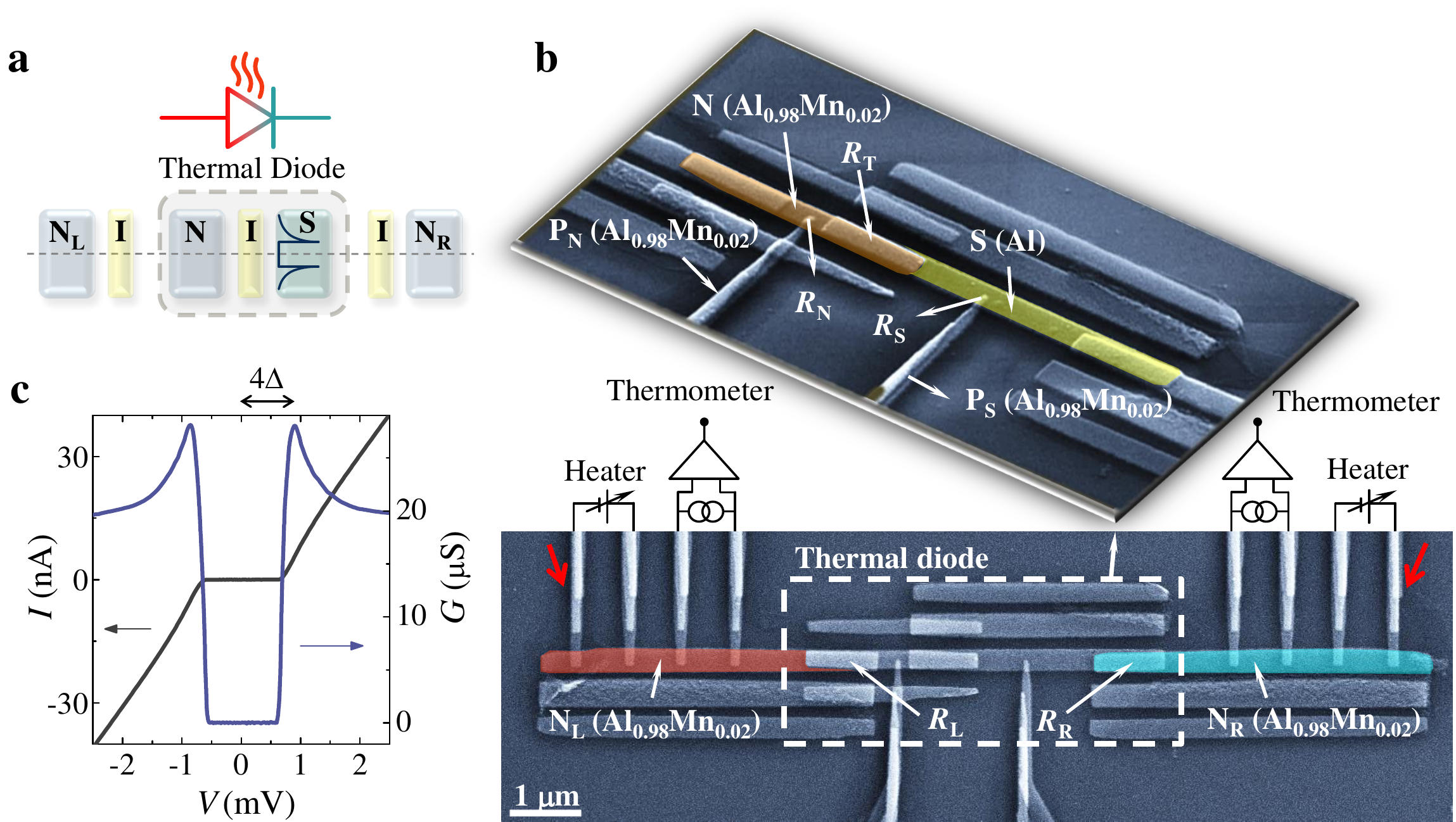}
\caption{\textbf{Thermal diode's implementation.} \textbf{(a)} The hybrid thermal diode is realized by means of a  N$_{\rm L}$ININ$_{\rm R}$ chain. \textbf{(b)} False-color scanning electron micrograph of the sample. Top panel shows the thermal diode's core made of Al (S), AlOx (I) and Al$_{0.98}$Mn$_{0.02}$ (N, P$_{\rm N}$ and P$_{\rm S}$). The bottom image shows the NIS junction inserted between two right and left Al$_{0.98}$Mn$_{0.02}$ electrodes which include four Al/AlOx wires ($\sim 20$ k$\Omega$ each) serving either as thermometers or heaters. $R_{\rm T}\approx 7.5$ k$\Omega$, $R_{\rm R}\approx 3.1$ k$\Omega$ and $R_{\rm L}\approx 1.8$ K$\Omega$ are the normal-state resistances of the NIS junction, right and left electrodes, respectively, whereas P$_{\rm N}$ and P$_{\rm S}$ exhibit normal-state resistances $R_{\rm N} \sim 43$ k$\Omega$ and $R_{\rm S} \sim 80$ k$\Omega$, respectively. \textbf{(c)} Experimental electric current ($I$, left axis) and conductance ($G$, right axis) characteristics vs. voltage ($V$) measured at $T_{\rm bath} = 50$ mK through the series connection of two superconducting heaters wires on N$_{\rm R}$ and N$_{\rm L}$ as indicated by the red arrows in the bottom of panel \textbf{b}. This leads to a total resistance $1/G\gtrsim50$ k$\Omega$ for bias voltage well above $4\Delta(0)/e$, where $\Delta(0) \approx230$ $\mu e$V is the zero-temperature superconducting energy gap of Al, and $e$ is the electron charge.}
\label{Fig1}
\end{figure*}

Here we experimentally demonstrate a thermal diode design which joins  the two aforementioned  strategies. In our device, thermal symmetry breaking is achieved thanks to the insertion of an S electrode in the normal metal diode, leading to a N$_{\rm L}$INISIN$_{\rm R}$ chain (see Fig. \ref{Fig1}a). At temperatures well below $T_{\rm c}$, this is equivalent to set $R_{\rm R}\gg R_{\rm L}$ since the presence of $\Delta$ suppresses drastically  the heat flow through the S electrode \cite{GiazottoRev}. The thermalizing cold finger \cite{FornieriAPL} is realized by means of a normal metal probe P$_{\rm N}$, which, owing to its large volume, is fully thermalized with bath phonons \cite{GiazottoRev,Wellstood}. Such a design requires a simpler fabrication protocol with respect to that analyzed in Ref. \cite{FornieriAPL}, and 
offers the possibility to explore both the rectification regimes mentioned above:
the diode does indeed efficiently rectify heat in the forward configuration and, as $T$ increases, enters the regime dominated by the $T$-dependence of $\Delta$ in which heat flows preferentially in the reverse configuration.

The thermal diode has been fabricated by electron beam lithography, three-angle shadow mask evaporation of metals and in-situ oxidation (see Methods Summary). Aluminum (Al) with $T_{\rm c} \approx 1.5$ K implements all superconducting parts of the structure whereas Al$_{0.98}$Mn$_{0.02}$ has been used as a normal metal \cite{MartinezNature,Maasilta}. The diode's core, enlarged in the top part of Fig. \ref{Fig1}b, consists of a NIS junction with normal-state resistance $R_{\rm T}$. 
Two normal metal probes, P$_{\rm N}$ and P$_{\rm S}$, are tunnel-coupled to the N and S electrodes through junctions with normal-state resistances $R_{\rm N}$ and $R_{\rm S}$, respectively. 
Apart from the crucial role of P$_{\rm N}$ in thermally coupling the N island to the bath,
these probes are also fundamental for the electrical characterization of the device. 
The whole structure is shown in the bottom part of Fig. \ref{Fig1}b. The N$_{\rm L}$ and N$_{\rm R}$ electrodes include four tunnel-coupled superconducting probes operating either as heaters or thermometers \cite{GiazottoRev,GiazottoNature,MartinezNature}. 
Four- and two-wire electric measurements performed through these probes together with P$_{\rm N}$ and P$_{\rm S}$ allow us to determine the normal-state resistance of all junctions present in the structure. 
Despite its intrinsic asymmetry, our device is fully symmetric from the \textit{electrical} side. This is confirmed by the differential conductance ($G=\partial I/\partial V$) obtained from the experimental current ($I$) vs. voltage ($V$) characteristic shown in Fig. \ref{Fig1}c (see Supplementary Materials for details).
  
\begin{figure}[h]
\centering
\includegraphics[width=0.65\columnwidth]{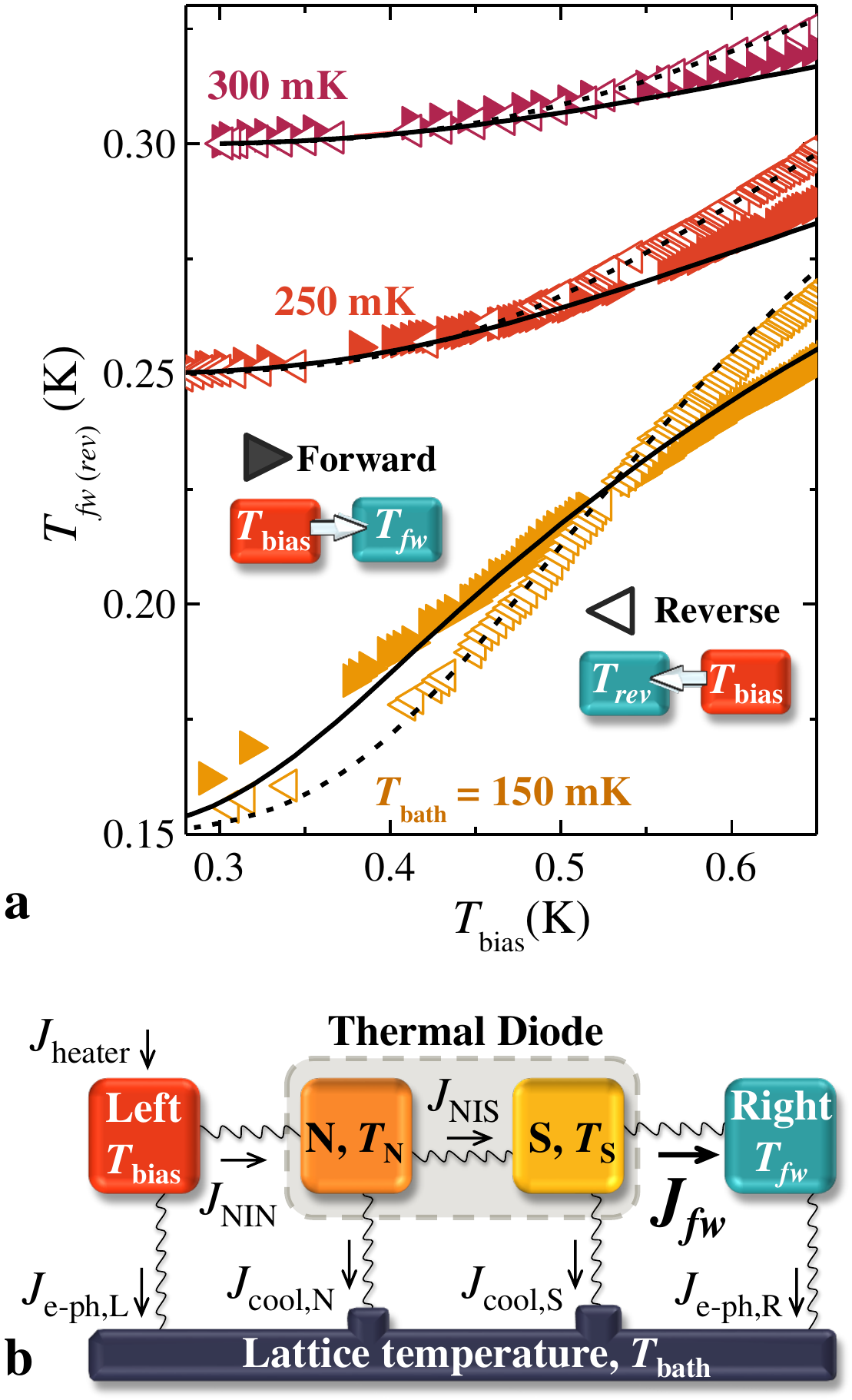}
\caption{\textbf{Thermal diode's response and modelling.} \textbf{(a)} Experimental electron temperature $T_{fw(rev)}$ vs.  $T_{\rm bias}$ measured at three representative bath temperatures $T_{\rm bath}$ and corresponding to the forward (full symbols) and reverse (open symbols) configurations, as illustrated in the legend. The experimental error lies within the size of the symbols. 
The thermal rectifying character of the diode is clearly reflected in the differences between $T_{fw}$ and $T_{rev}$.
Lines are the theoretical results obtained from the thermal model described below. \textbf{(b)} Thermal model outlining the relevant heat exchange mechanisms present in the structure. Arrows indicate the heat current directions in the forward configuration for $T_{\rm bias} > T_{\rm N} > T_{\rm S} > T_{fw} > T_{\rm bath}$.} 
\label{Fig2}
\end{figure}

By contrast, transport properties are highly asymmetric from the \textit{thermal} side. We stress that we are concerned with the heat current carried by electrons only. We indeed assume that lattice phonons present in the whole structure are thermalized with substrate phonons residing at the bath temperature ($ T_{\rm bath}$) and, consequently, are not responsible of any heat transport. 
This assumption is expected to hold as Kapitza resistance between thin metallic films and the substrate is vanishing at low temperatures \cite{Wellstood,GiazottoNature,MartinezNature}. Directional thermal current mismatch is demonstrated by $T$-biasing the structure so to create a thermal gradient across it. 
Each electrode of the device can therefore be described by a Fermi-like energy distribution characterized by an electronic temperature that can largely differ from $T_{\rm bath}$ \cite{Wellstood}. 
This is possible since electrons in micrometer-sized metallic electrodes  are weakly coupled to lattice phonons at sub-kelvin temperatures.
In the forward configuration, the bias temperature on N$_{\rm L}$ ($T_{\rm bias}$) is raised above $T_{\rm bath}$ while monitoring the resulting electronic temperature on N$_{\rm R}$ ($T_{fw}$). 
The measurement procedure is inverted in the reverse configuration, in which the temperature of N$_{\rm L}$ ($T_{rev}$) is probed for an increasing $T_{\rm bias}$ set in N$_{\rm R}$. 
Heating in N$_{\rm L}$ (and N$_{\rm R}$) is achieved by injecting Joule power through a couple of tunnel junctions whereas the electronic temperature is determined from the $T$-dependent voltage drop across the other couple of current-biased superconducting junctions \cite{GiazottoRev,GiazottoNature,MartinezNature}.

Three representative curves, obtained at different $T_{\rm bath}$, are shown in Fig. \ref{Fig2}a. Differences between the temperatures measured in the forward (full symbols) and reverse configurations (open symbols) 
are evident, becoming even more apparent by lowering $T_{\rm bath}$. 
These differences are, by themselves, proof of the thermal rectifying nature of the N$_{\rm L}$INISIN$_{\rm R}$ chain.
Furthermore, the intersection between $T_{fw}$ and $T_{rev}$, clearly visible at $T_{\rm bath}=150$ mK, 
pinpoints the crossover between the two rectification regimes that we anticipated above.

Since any direct measurement of the heat current is infeasible, the magnitude of heat rectification is assessed from the experimental temperatures with the aid of the thermal model sketched in Fig. \ref{Fig2}b describing the forward $T$-bias configuration. The four electrodes forming the device reside at temperatures $T_{\rm bias}>T_{\rm N}>T_{\rm S}>T_{fw}>T_{\rm bath}$, where $T_{\rm N}$ and $T_{\rm S}$ are the electronic temperatures of N and S, respectively. Terms $J_{\rm NIN}$ and $ J_{\rm NIS}$ represent the heat currents entering the thermal diode and flowing from N to S, respectively. $J_{\rm e-ph}$ terms account for heat exchanged between electrons and lattice phonons. In addition to this contribution, $J_{\rm cool}$ terms include also the energy losses trough P$_{\rm N}$ and P$_{\rm S}$ (see Methods Summary for details).

The steady-state electronic temperatures $T_{\rm N}$, $T_{\rm S}$ and $T_{fw}$ can be calculated for any given  $T_{\rm bias}$ and $T_{\rm bath}$ by solving the following system of energy-balance equations:
\begin{eqnarray}
J_{\rm NIN} (T_{\rm bias},T_{\rm N} )- J_{\rm cool,N}(T_{\rm N},T_{\rm bath}) - J_{\rm NIS}(T_{\rm N},T_{\rm S})  &= & 0,  \\
J_{\rm NIS}(T_{\rm N},T_{\rm S}) - J_{\rm cool,S} (T_{\rm S},T_{\rm bath})  - J_{fw} (T_{\rm S},T_{fw})  &=& 0, \\ 
J_{fw} (T_{\rm S},T_{fw}) - J_{\rm e-ph, R} (T_{fw},T_{\rm bath}) &=& 0.
\label{system}
\end{eqnarray}
In particular, Eqs. (1), (2) and (3) account for the detailed thermal budget in the N, S and right  electrode, respectively, by setting to zero the sum of all the incoming and outgoing heat currents. An analogous set of equations can be written as well for the reverse thermal bias configuration (see Methods Summary).

\begin{figure*}[t]
\centering
\includegraphics[width=0.7\textwidth]{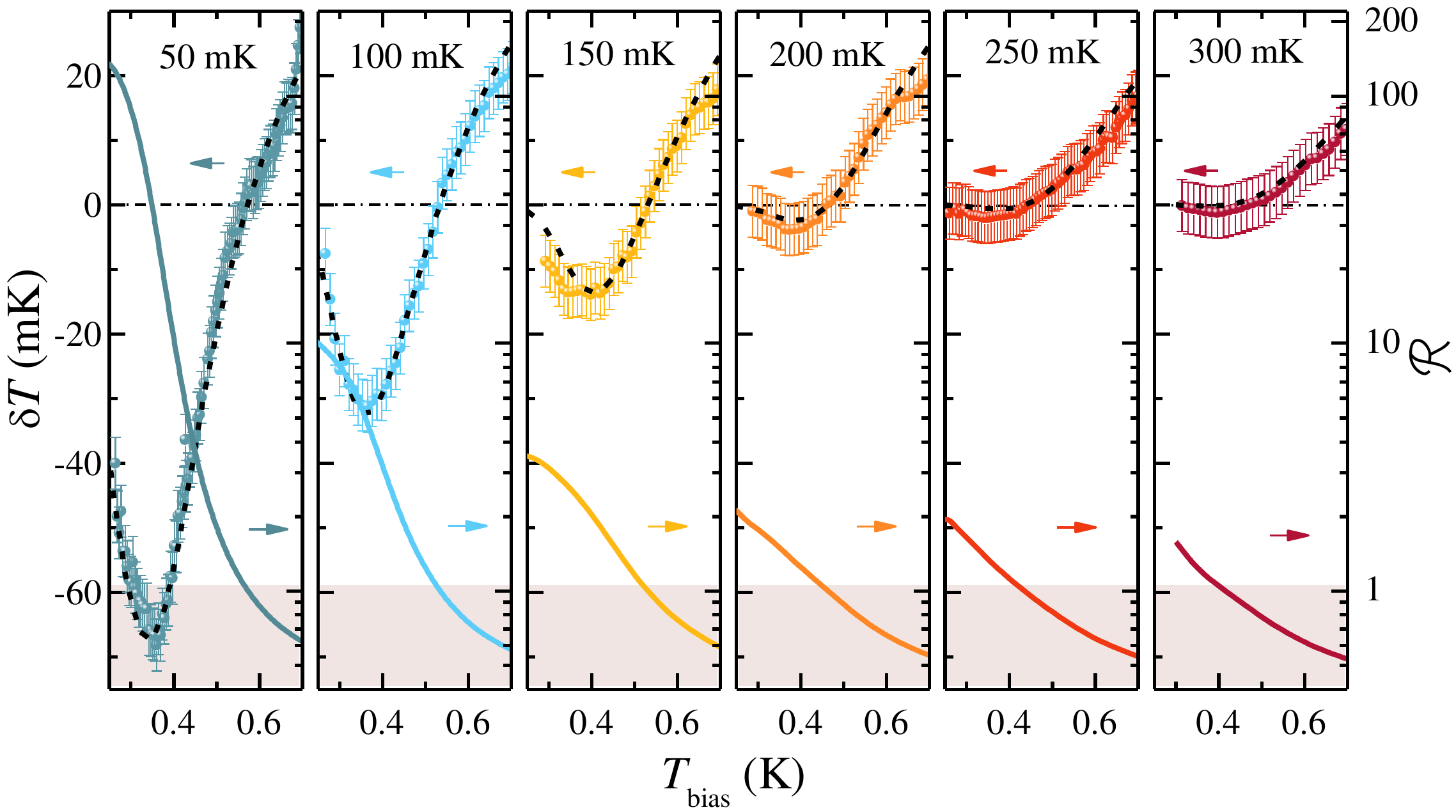}
\caption{\textbf{Thermal diode's performance.} Experimental temperature difference $\delta T = T_{rev}-T_{fw}$ vs.  $T_{\rm bias}$ (scatter, left axes) measured at different $T_{\rm bath}$ indicated at the top of each panel.  Error bar size is $\pm 4$ mK, and horizontal dash-dotted lines correspond to  $\delta T = 0$. 
Black dashed lines are the theoretical results from the thermal model. Colored lines are the corresponding thermal rectification coefficients $\mathcal{R}$  (right axes on a logarithmic scale). Shadowed regions evidence the regime in which $\mathcal{R} < 1$.}
\label{Fig3}
\end{figure*} 

The calculated $T_{fw}$ and $T_{rev}$ values are fitted to the 
measured data using the structure parameters determined experimentally. As only fitting parameter we have let $R_{\rm N}$ vary from 50\% to 100\% of its nominal value which accounts for additional energy losses from the diode's core due to non-idealities of the P$_{\rm N}$ junction (see Methods Summary). As shown in Fig. \ref{Fig2}a a good agreement between theory and experiment is obtained. This supports the validity of the thermal model which represents an essential tool to extract the value of $\mathcal{R}$.

Figure \ref{Fig3} summarizes our main results. For each $T_{\rm{bath}}$, the experimental temperature difference $\delta T=T_{rev}-T_{fw}$  is plotted vs. $T_{\rm{bias}}$ along with the theoretical curves obtained from the thermal model. As before, agreement between theory and experiment is noteworthy. The corresponding rectification ratio  $\mathcal{R}$  is plotted on the right axes. Large negative  $\delta T$ leads to $\mathcal{R} \gg 1$ stemming from substantial thermal rectification in the forward configuration. More specifically $|\delta T|$ reaches values exceeding $60$ mK for $T_{\rm{bias}}\sim 350$ mK and $\mathcal{R} \sim 140$ can be attained at the lowest bath temperature. As the bias temperature is raised, $|\delta T|$ decreases and $\mathcal{R}$ approaches unity indicating a reduction of the diode's performance. Further increase of $T_{\rm{bias}}$ leads to $\mathcal{R}<1$ pinpointing the outset of the other regime of rectification. In this case, however, differences between the forward and reverse heat currents are less pronounced leading only to  $\mathcal{R}\sim 0.5$ at $T_{\rm{bath}}=300$ mK.

\begin{figure}[b]
\includegraphics[width=0.9\columnwidth]{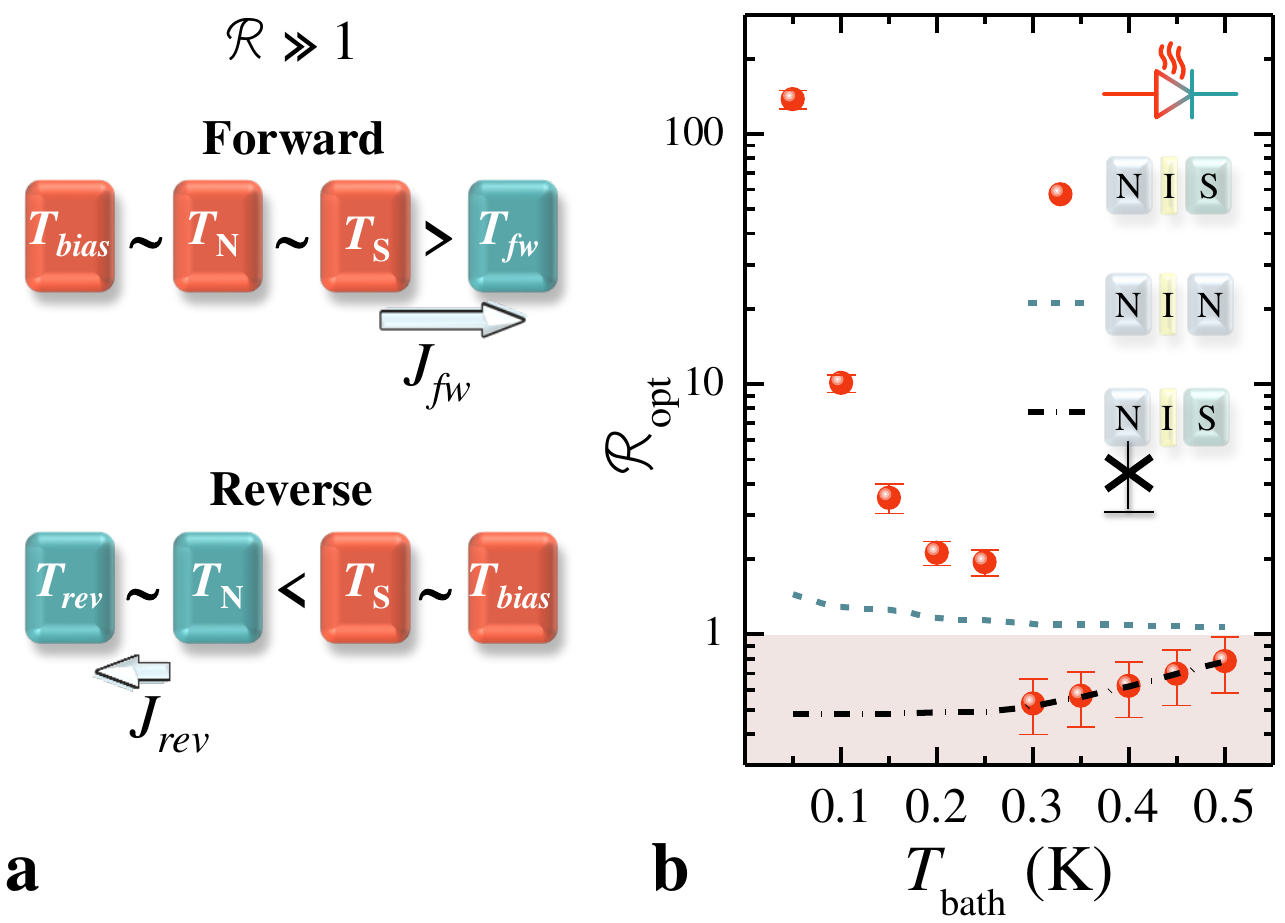}
\caption{\textbf{Heat rectification mechanisms.} \textbf{(a)} Pictorial representation of the diode's performance in both configurations when $\mathcal{R}\gg 1$. Red and blue colors indicate hot and cold temperatures, respectively. \textbf{(b)} Optimal efficiency $\mathcal{R}_{\rm{opt}}$ vs. $T_{\rm {bath}}$ obtained for the N$_{\rm{L}}$INISIN$_{\rm{R}}$ thermal diode (full circles) and for two additional cases of interest: the dashed line is calculated by replacing S with a normal metal whereas the dash-dotted line stands for the optimal efficiency obtained by removing P$_{\rm{N}}$. Shadowed region emphasizes the regime where $\mathcal{R} < 1$.}
\label{Fig4}
\end{figure}

We now focus on the mechanisms at the origin of the observed large forward thermal rectification, i.e., for $\mathcal{R}\gg 1$. The S electrode behaves as a bottleneck for the electronic heat current since quasiparticles with energy smaller than $ \Delta$ cannot tunnel through the forbidden energy gap. As pictorially sketched in Fig. \ref{Fig4}a, the bottleneck effect favors the development of large $T$-gradients in the forward configuration ($T_{\rm S}>T_{fw}$) but prevents $T_{\rm N}$ to rise considerably above $T_{\rm bath}$ in the reverse one. In the latter case, additionally, a large amount of the total heat current is released to the phonon bath through the cold finger probe P$_{\rm N}$ further lowering $T_{\rm N}$ down to $T_{\rm bath}$. The $T$-gradient developed in the reverse configuration ($T_{\rm N} \sim T_{rev}$) is therefore strongly reduced leading, finally, to $J_{fw}\gg J_{rev}$. 
We emphasize that both the peculiar temperature-dependence of the energy release through the contact ($\propto T^n$, with $n=2$) and the magnitude of 
the tunneling resistance $R_{\rm N}$ play a determinant role in the behavior of the thermal diode (see Methods Summary). As a matter of fact, the lower the value of $n$, the larger is the effectiveness of the cold finger at temperatures $<1$ K (to this end, a negative $n$ value would be ideal). Furthermore, according to our analysis, the experimental value of $R_{\rm N}$ is close to the one that maximizes the rectification ratio (see Supplemetary Materials). By contrast, the influence of P$_{\rm S}$ can be neglected due to the limited thermal conductance of the NIS contact \cite{GiazottoRev}. 

The aforementioned picture does not longer hold as  $T_{\rm bias}$ increases, when the device enters the regime of reverse thermal rectification, i.e., $\mathcal{R}<1$. We recall that the energy diverted from N to the bath, i.e., $J_{\rm cool,N}$, consists of two terms: the heat current flowing through the cold finger probe P$_{\rm N}$ ($\propto T^2$) and the heat exchanged with lattice phonons ($\propto T^6$). At high $T_{\rm bias}$, the $T^6$ term can eventually dominate over the $T^2$ contribution entailing a strong enhancement of the electron-phonon coupling in the N electrode. This prevents the development of large thermal gradients across the device in the forward configuration leading to a decrease of  $J_{fw}$ and $\mathcal{R}<1$. This scenario is intensified as $T_{\rm bath}$ increases and, for this reason, the transition to the reverse rectification regime occurs for decreasing $T_{\rm bias}$. Additionally, in the reverse configuration, the S electrode reaches temperatures large enough to affect the value of $\Delta(T_{\rm S})$. As predicted in Refs. \cite{MartinezAPLrect,GiazottoBergeret}, reverse thermal rectification is then further enhanced becoming even more efficient than the forward one at $T_{\rm bath}= 300$ mK (notice that $1/\mathcal{R}\sim 2$ at $T_{\rm bias}= 700$ mK whereas $\mathcal{R}\sim 1.5$ at $T_{\rm bias}=300$ mK).

This behavior is summarized in Fig. \ref{Fig4}b where the optimal rectification ratio ($\mathcal{R}_{\rm opt}$) is plotted against  $T_{\rm bath}$ (full circles). $\mathcal{R}_{\rm opt}$ is defined as that corresponding to the maximum value between $1/\mathcal{R}$ and $\mathcal{R}$. As previously mentioned, the diode exhibits a sudden change of behavior at $T_{\rm bath}= 300$ mK when $\mathcal{R}_{\rm opt}<1$. This transition temperature results from the competition between the $T^2$ and the $T^6$ terms and, therefore, can be modified by choosing different structural parameters such as the material forming the N electrode or the value of $R_{\rm N}$ (see Methods). 
We also note the vanishing of heat rectification ($\mathcal{R}_{\rm opt}\sim 1$) for $T_{\rm bath}$ approaching $500$ mK. This stems from increased electron-phonon coupling in the N elements of the diode which levels temperature gradients across the whole device in both configurations.

We conclude by focusing on the crucial role embodied by the S island and 
the probe P$_{\rm N}$. 
Our experimental results are compared to the expected behavior of the diode when the S electrode is replaced with a normal metal (dashed line). This has dramatic consequences on the efficiency leading only to a minute $\mathcal{R}_{\rm opt}\lesssim 1.5$. On the other side, if P$_{\rm N}$ is removed from the device, its response (dash-dotted line) changes drastically exhibiting $\mathcal{R}_{\rm opt}< 1$. Furthermore, this  line accounts well for the experiment for $T_{\rm bath}\geq 300$ mK. This suggests a reduction of the effectiveness of the probe P$_{\rm N}$ at higher $T_{\rm bath}$, and supports the proposed physical picture for the regime  $\mathcal{R}<1$.

In summary, we have realized a competitive hybrid thermal diode which provides highly efficient rectification of the electronic heat current (up to $\mathcal{R} \sim 140$). In addition, the  device allowed us to prove and investigate two different regimes of rectification that were recently predicted to occur \cite{MartinezAPLrect,GiazottoBergeret,FornieriAPL}.
The structure design, based on a simple N$_{\rm L}$INISIN$_{\rm R}$ chain, is easily implementable with conventional nanofabrication techniques and could be combined with, for instance, electronic coolers or radiation sensors \cite{GiazottoRev}. The thermal diode would enable heat evacuation and electrical control over such devices while limiting the power delivered into them. This technology might also have a potential impact in general-purpose cryogenic electronic microcircuitry, e.g., solid-state quantum information architectures \cite{NielsenChuang}. Combined with the heat interferometers reported in Refs. \citenum{GiazottoNature} and \citenum{MartinezNature}, our thermal rectifier materializes one of the main building blocks of forthcoming coherent caloritronic nanocircuits \cite{MartinezRev}.



\section*{Methods summary}

The samples were fabricated through electron-beam lithography and three-angle shadow-mask evaporation of metals onto an oxidized Si wafer through a suspended resist mask. In the electron-beam evaporator, the chip was initially tilted at an angle of 26$\mathrm{^{\circ}}$: a 20-nm-thick layer of Al was deposited to form the heater/thermometer probes along with the superconducting (S) part of the thermal diode. The samples were then exposed to 200 mTorr of O$\mathrm{_{2}}$ for 5 min to form the thin insulating layer of AlOx (I) in the heater/thermometer tunnel junctions, after which it was tilted to -26$^\mathrm{{\circ}}$ for the deposition of 25 nm of Al$_{0.98}$Mn$_{0.02}$ to form N$_{\rm R}$, N$_{\rm L}$, P$_{\rm N}$ and P$_{\rm S}$. The chip was subsequently exposed to 200 mTorr of O$\mathrm{_{2}}$ for 5 min to form the AlOx layer of the NIS tunnel junction at the core of the device. Finally, a 30-nm-thick layer of Al$_{0.98}$Mn$_{0.02}$ was deposited at 0$\mathrm{^{\circ}}$ to create the N part of the thermal diode. The N$_{\rm R}$ and N$_{\rm L}$ electrodes are nominally identical, and have a volume $\mathcal{V}\mathrm{_{L,R}}$ = 2.3 x 10$^{-20}$ m$^3$. On the other hand, each section of the thermal diode has a volume $\mathcal{V}\mathrm{_{N,S}}$ = 1.6 x 10$^{-20}$ m$^3$. 

The electric characterization of the sample was performed down to 50 mK in a filtered dilution refrigerator. Current biasing of the thermometers was obtained through battery-powered floating sources, whereas the heaters were operated upon voltage biasing within 0-2 mV, corresponding to a maximum of $\sim$ 50 pW of power injected into the N electrodes. Thermometer bias currents were varied from 5 pA to 100 pA in order to achieve high sensitivity in different ranges of temperature while limiting the impact of self-heating and self-cooling \cite{GiazottoRev}. Voltage and current were measured with conventional room-temperature preamplifiers.

In the thermal model, $J_{fw}(T_\mathrm{S},T_\mathrm{fw})=J_\mathrm{NIS}(T_\mathrm{S},T_{fw})$ where $J_\mathrm{NIS}(T_{\rm 1},T_{\rm 2})=\frac{2}{e^2 R_\Omega} \int_0^{\infty} dE E$ 
$\mathcal{N}(E,T_{\rm 1})  [f(E,T_{\rm 1})-f(E,T_{\rm 2})]$ \cite{GiazottoRev}, $f(E,T)=[1+\textrm{exp}(\frac{E}{k_{\rm B}T})]^{-1}$ is the Fermi-Dirac distribution function, $ \mathcal{N}(E , T)=\left| \Re [  E+i \Gamma/ \sqrt{(E+i \Gamma)^2- \Delta^2(T)} ] \right|$ is the smeared (by non-zero $\Gamma$) normalized Bardeen-Cooper-Schrieffer density of states in the superconductor \cite{Dynes}, $R_\Omega$ is the tunnel junction normal-state resistance, $e$ is the electron charge, and $k_B$ is the Boltzmann constant. On the other side, $J_{\rm NIN}(T_{\rm 1},T_{\rm 2})=\frac{\pi^2 k_{\rm{B}}^2}{6e^2 R_\Omega}(T_{\rm 1}^2-T_{\rm 2}^2)$ \cite{GiazottoBergeret}, $J_{\rm cool,N}(T_{\rm N},T_{\rm bath})= J_{\rm NIN}(T_{\rm N},T_{\rm bath})+J_{\rm e-ph,N}(T_{\rm N},T_{\rm bath})$, and  $J_{\rm cool,S}(T_{\rm S},T_{\rm bath})=  J_{\rm NIS}(T_{\rm S},T_{\rm bath})+J_{\rm e-ph,S}(T_{\rm S},T_{\rm bath})$.
Above, $J_{\rm{e-ph,N}}(T,T_{\rm bath})=\Sigma_{\rm N} \mathcal{V}_\mathrm{N}(T^{n}-T_{\rm bath}^{n})$ and $J_{\rm{e-ph,S}}(T,T_{\rm bath})=-\frac{\Sigma_{\rm S} \mathcal{V}_{\rm S}}{96 \zeta (5)k_{\rm B}^5}\int_{-\infty}^{\infty} dE E \times $ \linebreak $ \int_{-\infty}^{\infty} d\rm{\epsilon}\rm{\epsilon}^2 \rm{sgn}(\rm{\epsilon})L_{E,E+\rm{\epsilon}}\big[\rm{coth}(\frac{\rm{\epsilon}}{2k_{\rm B}T_{\rm bath}})(f_E-f_{E+\rm{\epsilon}})-f_E f_{E+\rm{\epsilon}}+1 \big]$ \cite{Timofeev}. Here, $\Sigma_{\rm N,S}$ is the material-dependent electron-phonon coupling constant and $n$ is the characteristic exponent of the material. Additionally, $f_E=\tanh (\frac{E}{2k_B T})$ and $L_{E,E'}=\mathcal{N}(E,T)\mathcal{N}(E',T)[1-\frac{\rm{\Delta^2(T)}}{EE'} ]$. In our case $n=$ 6, $\Sigma_{\rm AlMn}=$ 4.5 $\times$ 10$^9$ WK$^{-6}$m$^{-3}$ \cite{Maasilta,MartinezNature} and $\Sigma_{\rm Al}=$ 0.3 $\times$ 10$^9$ WK$^{-5}$m$^{-3}$ \cite{GiazottoRev} corresponding to  Al$_{0.98}$Mn$_{0.02}$ and Al, respectively. On the other hand, in the reverse $T$-bias configuration the energy-balance equations read $J_{\rm NIS}(T_{\rm bias},T_{\rm S})-J_{\rm NIS}(T_{\rm S},T_{\rm N})-J_{\rm cool,S}(T_{\rm S},T_{\rm bath})=0$, $J_{\rm NIS}(T_{\rm S},T_{\rm N})-J_{rev}(T_{\rm N},T_{rev})-J_{\rm cool,N}(T_{\rm N},T_{\rm bath})=0$ and $J_{rev}(T_{\rm N},T_{rev})-J_{\rm e-ph,L}(T_{rev},T_{\rm bath})=0$ where $J_{rev}(T_{\rm N},T_{\rm rev})=J_{\rm NIN}(T_{\rm N},T_{rev})$.  

Experimental data have been fitted by setting the measured values of $R_{L}$, $R_{T}$, $R_{S}$, $R_{R}$, $\Delta(0)$, and $\Gamma =10^{-4}\Delta(0)$ as determined from the electrical characterization of the device \cite{pekolaprl}. $R_{\rm{N}}$ is the only fitting parameter that was set to $ \sim 0.5\times R_{\rm{N}}^{exp}$ for $T_{\rm{bath}}\leq 250$ mK, 
$R_{\rm{N}}^{exp}$ being the measured value. We attribute this behavior to possible non-idealities of the Al$_{0.98}$Mn$_{0.02}$ oxidized tunnel junction, e.g., pinholes, that provide additional channels for heat transport leading to an effective reduction of $R_{\rm{N}}$. 
For  $T_{\rm{bath}}\geq 300$ mK, i.e., when the N probe does not longer play a role, data are well accounted for the theoretical lines when setting $R_{\rm{N}} = R_{\rm{N}}^{exp}$. 
Photon-mediated  thermal transfer between N$_{\rm L}$ and N$_{\rm R}$ has been neglected due to  poor impedance matching between them \cite{Pascal,meschke,schmidt}. 
Finally, we also checked numerically that distinct phononic temperatures that might exist in the different structure elements thanks to a finite Kapitza resistance affect insignificantly the results.

\section*{Supplementary material}
\subsection{Designing an efficient thermal diode}
\begin{figure*}[t]
\centering
\includegraphics[width=0.8\textwidth]{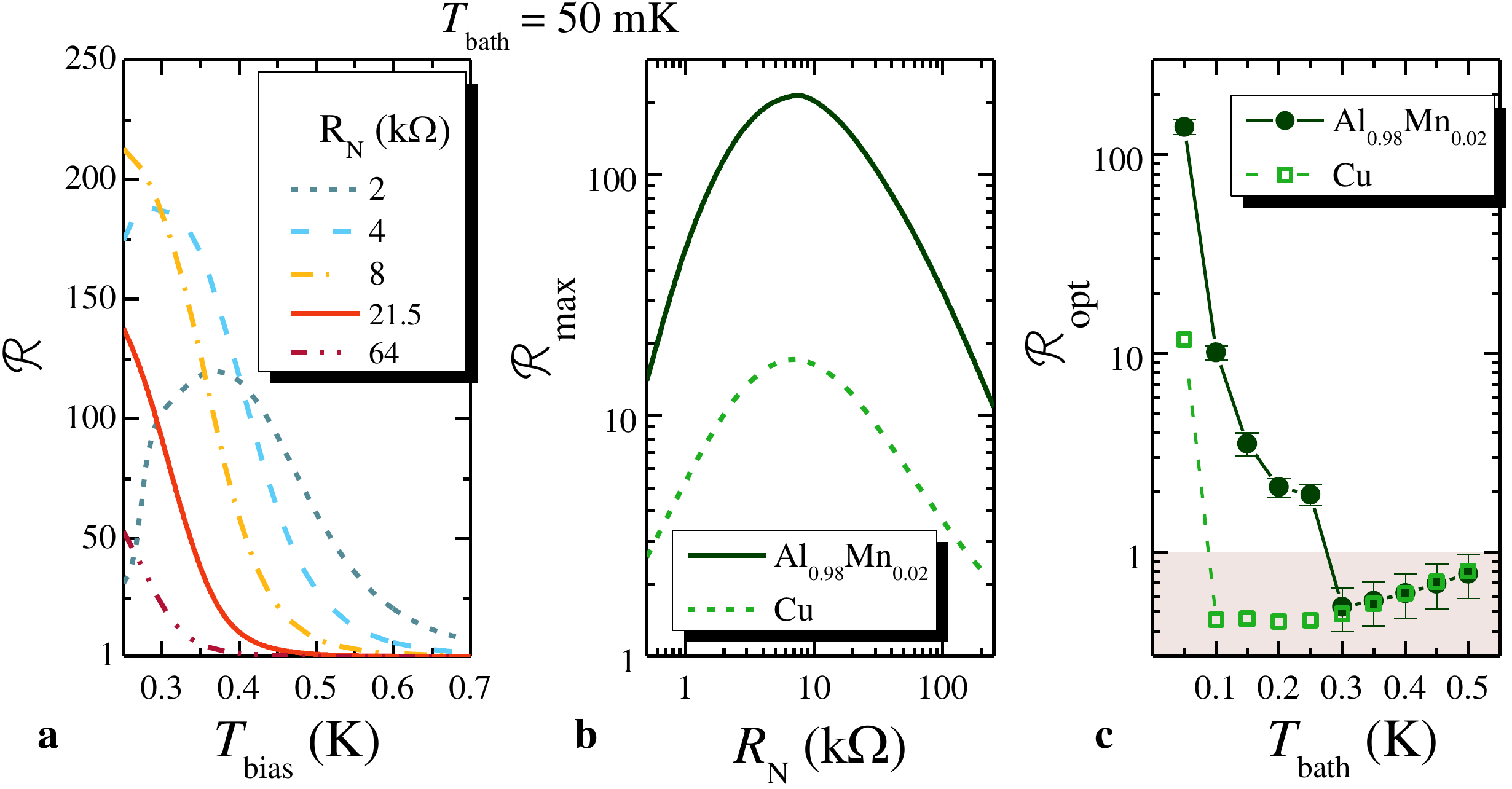}
\caption{\textbf{Role of the coupling strength to the phonon bath.} \textbf{(a)} Rectification efficiency $\mathcal{R}$ vs. $T_\mathrm{bias}$ calculated for different values of the probe resistance $R_\mathrm{N}$ at $T_\mathrm{bath}=$ 50 mK. The solid line is the result corresponding to our device with $R_\mathrm{N}=$ 21.5 k$\Omega$, as obtained from the theoretical fit of the experimental data. \textbf{(b)} Maximum rectification efficiency ($\mathcal{R}_\mathrm{max}$) vs. $R_\mathrm{N}$ calculated for two possible choices of the N material in the device at $T_\mathrm{bath}=$ 50 mK. \textbf{(c)} Optimal rectification efficiency $\mathcal{R}_\mathrm{opt}$ vs. $T_\mathrm{bath}$ for the same cases analyzed in panel B. Lines are guides to the eye. The curves have been calculated by setting $R_\mathrm{N}$ equal to the values obtained from the experimental fitting at each $T_\mathrm{bath}$. Shadowed region indicates the regime where $\mathcal{R} < 1$.}
\label{Fig1_Supp}
\end{figure*} 
As discussed in the main text, an efficient thermal diode can be realized by means of a N$_{\rm L}$INISIN$_{\rm R}$ chain, in which the N electrode is coupled to the phonon bath. This can be efficiently realized through a thermalizing normal metal probe P$_\mathrm{N}$ \cite{FornieriAPL}. As shown in Fig. \ref{Fig1_Supp}a, the value of the resistance $R_\mathrm{N}$ has a strong influence on the behavior of the rectification efficiency $\mathcal{R}$ vs. the bias temperature $T_\mathrm{bias}$: not only it modifies the maximum value of $\mathcal{R}$ ($\mathcal{R}_{\rm max}$), but it also moves the maximum point. This is strictly related to the diode working principle: at low temperatures the S electrode works as a thermal bottleneck in both the forward and reverse configuration. Under equal $T_\mathrm{bias}$, we obtain $\mathcal{R}\gg 1$ if the N part of the diode is efficiently thermalized to the bath temperature $T_{\rm bath}$, and $R_\mathrm{N}$ plays an important role in this process. As a matter of fact, if P$_{\rm N}$ is too transparent, $T_\mathrm{N}$ and $T_\mathrm{S}$ in the forward configuration become too close to $T_{fw}$; on the contrary, if P$_{\rm N}$ is too opaque, in the reverse configuration $T_\mathrm{N}$ gets significantly higher than $T_{rev}$. This delicate trade-off leads to a non-monotonic behavior of $\mathcal{R}_{\rm max}$ vs. $R_\mathrm{N}$, as shown in Fig. \ref{Fig1_Supp}b.

The role of $R_\mathrm{T}$, on the other side, turns out not to be so crucial to optimize the efficiency of the heat diode. Increasing or decreasing $R_\mathrm{T}$ leads, respectively, to the enhancement or reduction of the thermal gradients at the diode's output in both the configurations. This yields a non-monotonic behavior of $\mathcal{R}_\mathrm{max}$ vs. $R_\mathrm{T}$ that does not differ significantly from the result we obtained for our device ($\mathcal{R}_\mathrm{max}$ shows a relative variation of $\sim$10\% for $R_\mathrm{T}$ spanning from 1 k$\Omega$ to 300 k$\Omega$).

In order to design an efficient thermal diode, the ingredient that has to be also considered is $J_\mathrm{e-ph,N}$, i.e., the electron-phonon relaxation in the N electrodes. In our experiment we exploited Al$_{0.98}$Mn$_{0.02}$ because of its favorable oxidation properties and of the reduced electron-phonon coupling at low temperatures. As a matter of fact, this material follows a $T^6$ power law with $\Sigma_\mathrm{AlMn}=$ 4.5 $\times$ 10$^9$ WK$^{-6}$m$^{-3}$ \cite{Maasilta,MartinezNature}. Another material that is typically used to fabricate N electrodes is copper (Cu), which is characterized by a $T^5$ dependence and $\Sigma_\mathrm{Cu}=$ 3 $\times$ 10$^9$ WK$^{-5}$m$^{-3}$\cite{GiazottoRev,GiazottoNature}. 

In Fig. \ref{Fig1_Supp}b we compare the maximum rectification efficiency obtained at $T_{\rm bath} = 50$ mK as a function of $R_\mathrm{N}$ for two different choices of the N material. We limit our analysis to the same range of $T_\mathrm{bias}$ we explored experimentally. The solid lines correspond to a diode made of Al$_{0.98}$Mn$_{0.02}$ and Al, identical to the one we measured, whereas the dashed lines stand for a device where all the N parts are made of Cu. In the latter case, the $T^5$ dependence of the electron-phonon coupling has dramatic consequences on the efficiency of the device. This stems from the stronger energy relaxation that affects all the N electrodes, thereby levelling temperature gradients across the whole device.

Finally, we consider the role of the electron-phonon coupling for an increasing $T_\mathrm{bath}$. To this end, we define the optimal rectification ratio $\mathcal{R}_\mathrm{opt}$ as the one corresponding to the maximum value between $\mathcal{R}$ and $1/\mathcal{R}$. In Fig. \ref{Fig1_Supp}c we plot $\mathcal{R}_\mathrm{opt}$ vs. $T_\mathrm{bath}$ for the same cases analyzed in Fig. \ref{Fig1_Supp}b. As explained in the main text, the optimal efficiency of the Al$_{0.98}$Mn$_{0.02}$ diode changes direction ($\mathcal{R}_\mathrm{opt}<1$) \cite{MartinezAPLrect,GiazottoBergeret} at $T_\mathrm{bath}=300$ mK, indicating a strong reduction in the effectiveness of P$_{\rm N}$. This switching temperature results from the competition between the energy release through the probe ($\propto T^2$) and the electron-phonon coupling ($\propto T^6$) affecting all the N electrodes of the structure. In the case of Cu, the switching point occurs at lower $T_\mathrm{bath}$, since the electron-phonon coupling in this material is stronger at low temperatures ($\propto T^5$).

\subsection{Electrical vs. thermal behavior of the device}

The electrical behavior of the device is symmetric whereas the thermal one is not because of the subtle difference between the electric and heat current. Let us focus first on the NIS junction at the core of the thermal diode. Under thermal bias, the heat current ($J$) flowing between the S electrode at temperature $T_{\rm S}$ and the N electrode at temperature $T_{\rm N}$ connected by a tunnel barrier with normal-state resistance $R_{\rm \Omega}$ reads \cite{GiazottoRev}:
\begin{align}
J(T_{\rm S},T_{\rm N})=\frac{2}{e^2 R_{\rm \Omega}}&\int_{0}^{\infty} dE E \mathcal{N}(E,T_{\rm S})\\ \notag
&\times [f(E,T_{\rm S})-f(E,T_{\rm N})],
\end{align}
where $f(E,T)=[1+\textrm{exp}(\frac{E}{k_{\rm B}T})]^{-1}$ is the Fermi-Dirac distribution function and $\mathcal{N}(E , T)=\left| \Re [  E+i \Gamma/ \sqrt{(E+i \Gamma)^2- \Delta^2(T)} ] \right|$ is the smeared (by non-zero $\Gamma$) normalized Bardeen-Cooper-Schrieffer density of states in the superconductor \cite{Dynes}. On the other side, the electric current ($I$) flowing through the same junction under voltage bias $V$ can be written as follows \cite{GiazottoRev}:
\begin{align}
I(T_{\rm S},T_{\rm N},V)=\frac{1}{2 e R_{\rm \Omega}}&\int_{-\infty}^{\infty} dE \mathcal{N}(E,T_{\rm S})\\ \notag
&\times [f(E-eV,T_{\rm N})-f(E+eV,T_{\rm N})].
\end{align}
The peculiar dependence on temperature of the heat current $J$ makes it non-symmetric (i.e., without definite parity) under thermal bias reversal, i.e.,$|J(T_{\rm S},T_{\rm N})|  \neq |J(T_{\rm N},T_{\rm S})|$, which is at the origin of the thermal rectifying character of the NIS junction. By contrast, as it can be seen from the expression of the electric current, upon voltage bias reversal one obtains $|I(T_{\rm S},T_{\rm N},V)|=|I(T_{\rm S},T_{\rm N},-V)|$, which yields a symmetrical differential conductance. The electrical response of the whole device is the results of the series connection of additional NIN and NIS tunnel junctions, which all possess symmetric behavior from the electrical point of view.

If we focus into the whole diode, thermal asymmetry arises also as consequence of the asymmetric coupling to the phonon bath. This fact, again, does not affect the electric current flowing through the device. Indeed, the electric measurements shown in the main text were obtained through the series connection of two heater wires and at a uniform structure temperature equal to $T_{\rm bath}$, while letting the rest of wires electrically open. As a consequence, no electric current was flowing through the P$_{\rm N}$ probe in this case. 



\end{document}